\documentclass[11pt,a4paper,prb,nofootinbib,onecolumns,
superscriptaddress,amsmath,amssymb]{revtex4-1}
\usepackage{braket}
\usepackage{epstopdf,graphicx}
\usepackage{hyperref}
\hypersetup{%
colorlinks,
linkcolor={black},
citecolor={black},
urlcolor={black}}

\newcommand{\half}{\frac{1}{2}}
\newcommand{\roottwo}{\sqrt{2}}

\begin{document}

\title{Phase protection of Fano-Feshbach resonances}

\author{Alexander Blech}
\thanks{These authors have contributed equally.}
\affiliation{Theoretische Physik, Universit\"{a}t Kassel,
Heinrich-Plett-Stra{\ss}e 40, 34132 Kassel, Germany}

\author{Yuval Shagam}
\thanks{These authors have contributed equally.}
\altaffiliation[present address: ]{JILA, NIST and the Department of Physics, University of Colorado, Boulder CO 80309, USA}
\affiliation{Department of Chemical and Biological Physics, Weizmann Institute of Science, 
Rehovot 76100, Israel}

\author{Nicolas H\"olsch}
\altaffiliation[present address: ]{Laboratorium f\"ur Physikalische Chemie, 
ETH Z\"urich, 8093 Z\"urich, Switzerland}
\affiliation{Department of Chemical and Biological Physics, Weizmann Institute of Science, 
Rehovot 76100, Israel}

\author{Prerna Paliwal}
\affiliation{Department of Chemical and Biological Physics, Weizmann Institute of Science, 
Rehovot 76100, Israel}

\author{Wojciech Skomorowski}
\altaffiliation[present address: ]{Department of Chemistry, 
University of Southern California, Los Angeles, California 90089, USA}
\affiliation{Theoretische Physik, Universit\"{a}t Kassel,
Heinrich-Plett-Stra{\ss}e 40, 34132 Kassel, Germany}

\author{John W. Rosenberg}
\affiliation{Department of Chemical and Biological Physics, Weizmann Institute of Science, 
Rehovot 76100, Israel}

\author{Natan Bibelnik}
\affiliation{Department of Chemical and Biological Physics, Weizmann Institute of Science, 
Rehovot 76100, Israel}

\author{Oded Heber}
\affiliation{Department of Particle Physics and Astrophysics, 
Weizmann Institute of Science, 
Rehovot 76100, Israel}

\author{Daniel M. Reich}
\affiliation{Theoretische Physik, Universit\"{a}t Kassel,
Heinrich-Plett-Stra{\ss}e 40, 34132 Kassel, Germany}

\author{Edvardas Narevicius}
\affiliation{Department of Chemical and Biological Physics, Weizmann Institute of Science, 
Rehovot 76100, Israel}

\author{Christiane P. Koch}
\email[corresponding author: ]{christiane.koch@uni-kassel.de}
\affiliation{Theoretische Physik, Universit\"{a}t Kassel,
Heinrich-Plett-Stra{\ss}e 40, 34132 Kassel, Germany}
\affiliation{Department of Chemical and Biological Physics, Weizmann Institute of Science, 
Rehovot 76100, Israel}

\hyphenation{Fesh-bach}

\date{\today}

\maketitle


Decay of bound states due to coupling with free particle states is a general  phenomenon occurring at energy scales from MeV in nuclear physics to peV in ultracold atomic gases. 
Such a coupling gives rise to Fano-Feshbach resonances (FFR)\cite{FeshbachAnnPhys58,FanoPR61} 
that have become key to understanding and controlling interactions---in ultracold atomic gases,\cite{ChinRMP10} but also between quasiparticles such as microcavity polaritons.\cite{TakemuraNatPhys14}
The energy positions of FFR were shown to follow quantum chaotic statistics.\cite{FrischNature14,MaierPRX15} In contrast, lifetimes which are the fundamental property of a decaying state, have so far escaped a similarly comprehensive understanding. 
Here we show that a bound state, despite being resonantly coupled to a scattering state, becomes protected from decay whenever the relative phase is a multiple of $\pi$. We  observe this phenomenon by measuring lifetimes spanning four orders of magnitude for FFR of spin-orbit excited molecular ions 
with  merged beam and electrostatic trap experiments. Our results provide a blueprint for identifying naturally long-lived states in a decaying quantum
system.


Fano-Feshbach resonances describe decay of quantum mechanical bound states due to coupling with a continuum of scattering states 
which may be due to an effective nucleon-nucleon interaction in nuclear physics,\cite{FeshbachAnnPhys58} configuration interaction in autoionization,\cite{FanoPR61} spin-orbit interaction in rovibrational predissociation,\cite{CarringtonChemPhys85} or hyperfine interaction in ultracold gases.\cite{ChinRMP10}
A qualitative understanding of FFR lifetimes is obtained by first order perturbation theory, with Fermi's golden rule predicting them in terms of the coupling strength at resonance. Within this framework, the phenomenon of phase protection is readily understood  with just the most basic concepts of quantum theory. 

\begin{figure}[tb]
\centering
\includegraphics[width=\linewidth]{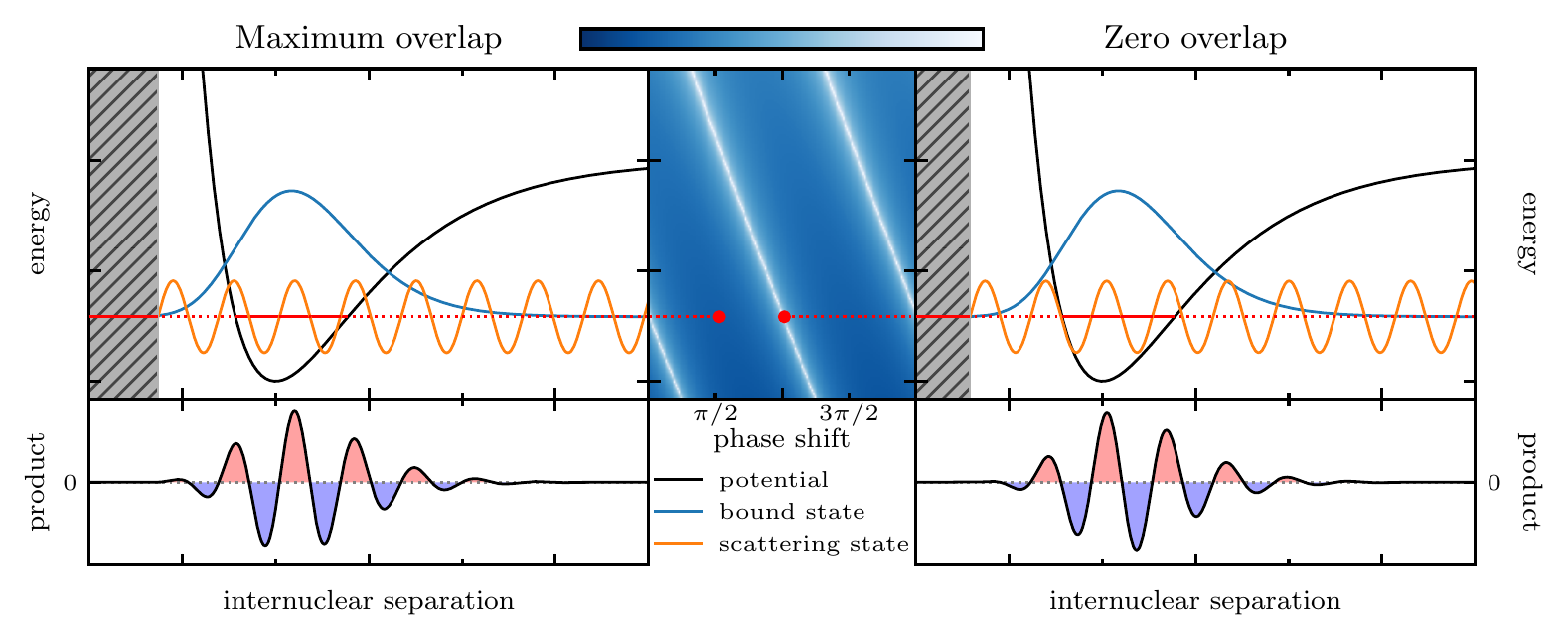}
\caption{The overlap of bound and continuum state determines, at least within the perturbative limit, the lifetime of an FFR: Vanishing overlap results in infinite lifetime, indicated by white lines in the center panel. The scattering phase is determined by the position of the scattering channel's repulsive wall, exemplified by the grey shaded area.}
\label{fig:scheme}
\end{figure}

Assuming the bound state to be an oscillator eigenstate weakly coupled to continuum states, and taking the coupling to be constant, the FFR lifetime is simply given by the overlap of the oscillator state $\ket{n}$ with the plane wave $\ket{k}$ that is resonant. This overlap is  the Fourier transform $\tilde\psi_n(k)$
of the $n$th oscillator eigenfunction. The lifetime of $\ket{n}$ becomes infinite when the overlap vanishes, i.e., at the roots $k^{(0)}_j$ of the Fourier transform, and provided there exist no other decay mechanisms. 
When accounting for the interaction between the colliding free particles, the plane waves are replaced by scattering 
functions of the form $\sin(kx+\delta)$ with 
$\delta$ the scattering phase shift.
Zero overlap with a bound state $\ket{n}$ 
then corresponds to the condition of vanishing imaginary part of 
$e^{i\delta}\tilde\psi_n(k)$, or, equivalently, 
\begin{equation}
\label{eq:condition}
\arg\left[\tilde\psi_n\left(k\right)\right] +\delta
=m\pi\quad \mathrm{with}\quad m\in \mathbb{Z}. 
\end{equation}
Equation~\eqref{eq:condition} implies that, 
for a given scattering momentum $k$, there exist phase shifts $\delta$ such that the complex overlap between bound state $\ket{n}$ and scattering state $\ket{k,\delta}$ vanishes. The lifetime of $\ket{n}$ may thus become infinite despite non-zero coupling strength.\footnote{Strictly speaking, the lifetime does not become infinite but will rather be limited by other decay processes such as radiative decay} This is illustrated in Fig.~\ref{fig:scheme}. 
While in a typical scattering event, $k$ and $\delta$ cannot be tuned independently, external field control modifies the phase and may thus indeed allow for extremely large lifetimes. To date, two rather different realizations of this phenomenon have been observed, respectively suggested --- magnetic field control of $s$-wave collisions in ultracold atomic gases~\cite{KoehlerPRL05} and electric dc field control of atomic photoionization.\cite{GilaryPRL08} 
While the required dc field strengths in the latter case are not yet available in experiment, magnetic field control over the lifetime of an FFR in ultracold collisions has been demonstrated experimentally.\cite{ThompsonPRL05} In that case, the magnetic field modifies the $s$-wave scattering length which is directly linked to the $s$-wave scattering phase shift.\cite{ChinRMP10}

\begin{figure}[tb]
\centering
\includegraphics[width=\linewidth]{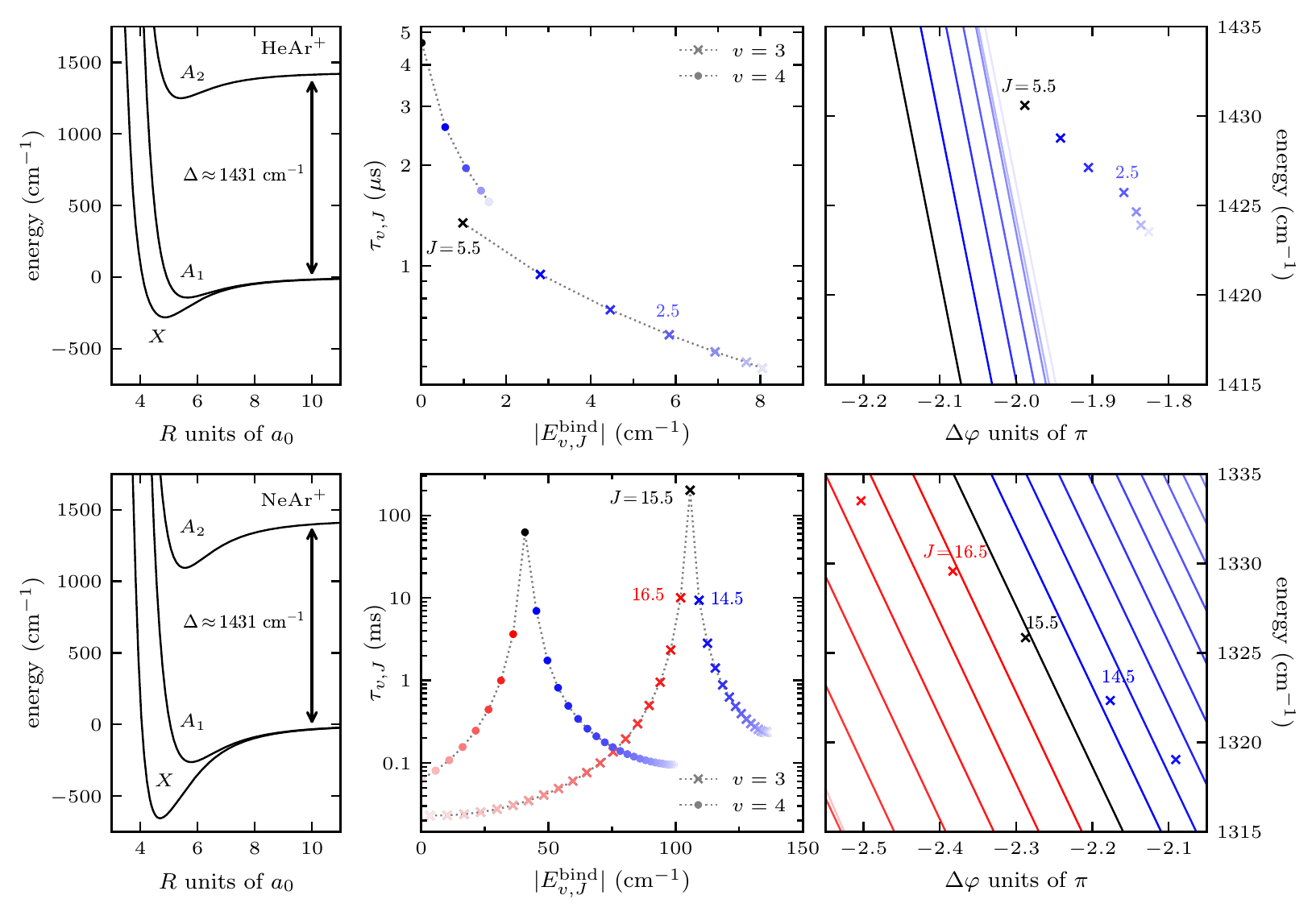} 
\caption{Rovibrational FFR for HeAr$^+$ (top) and NeAr$^+$ (bottom): Potential energy curves (left), lifetimes of $A_2$-state rovibrational levels, obtained with full account of non-perturbative effects, as a function of binding energy (center), and zero-overlap lines for fixed $v$ and different $J$ as a function of scattering phase and energy (right). 
}
\label{fig:lifetimes}
\end{figure}

Here, we show that the same effect can also be observed without external field control, just by ensuring that condition~\eqref{eq:condition} is fulfilled for certain bound states. 
Since $\hbar k=\sqrt{2\mu E}$, the number of roots of $e^{i\delta}\tilde\psi_n(k)$ as a function of energy increases with the reduced mass $\mu$. Thus, for a given energy $E$, 
the chance to fulfill Eq.~\eqref{eq:condition} increases as well.
We consider FFR in two different rare gas diatomic ions, HeAr$^+$ and NeAr$^+$. Rovibrational bound states in the $A_2$ potential energy curve,
cf. Fig.~\ref{fig:lifetimes} (left), with binding energy $E^{bind}_{v,J}$, 
have a finite lifetime due to the spin-orbit interaction which couples them to the scattering continuum of the electronic ground state $X$. While the coupling strength is identical for the two molecules, 
their reduced mass differs by a factor of 3.66, and NeAr$^+$ shows stronger binding in all of the relevant electronic states. As described in the Methods section, our theoretical model accounts non-perturbatively for relativistic and angular couplings. 
Phase protection is observed in the lifetimes' rotational progression for several vibrational states of NeAr$^+$. This is shown in Fig.~\ref{fig:lifetimes} (center) for two vibrational quantum numbers which are predominantly populated in the experiments discussed below. Their peak lifetimes differ from the shortest ones by four orders of magnitude. A similar behavior of lifetimes vs rotational energy has also been observed for HeNe$^+$.~\cite{CarringtonChemPhys85}
In contrast, phase protection does not play any role for HeAr$^+$ where lifetimes differ at most by a factor of 10. The right-hand side of Fig.~\ref{fig:lifetimes} explains this observation by comparing the distance of the actual energies and phases of the resonant scattering states (indicated by crosses) from the lines of vanishing overlap. Note that, for each rotational quantum number $J$, we get a plot as shown in the center of Fig.~\ref{fig:scheme}, due to the $J$-dependence of the rotational barrier. For ease of comparison, we only show the closest line of vanishing overlap in Fig.~\ref{fig:lifetimes}. We find that the closer a resonance is placed to its zero-overlap line, the larger becomes its lifetime. The difference between HeAr$^+$ and NeAr$^+$ is thus understood in terms of sampling in the phase-energy-plane. Approximating the $A_2$ state by a Morse oscillator, cf. Methods, we find the number of roots, i.e., the density of zero overlap lines, to increase with reduced mass as well as depth and equilibrium position of the potential and to decrease with potential width. This is in accordance with the \textit{ab initio} data for HeAr$^+$ and NeAr$^+$, cf. Fig.~\ref{fig:lifetimes}, since NeAr$^+$ possesses the larger reduced mass and deeper $A_2$ state potential whereas width and equilibrium distance are very similar for the two molecules.

\begin{figure}[tbp]
\centering
\includegraphics[width=\linewidth]{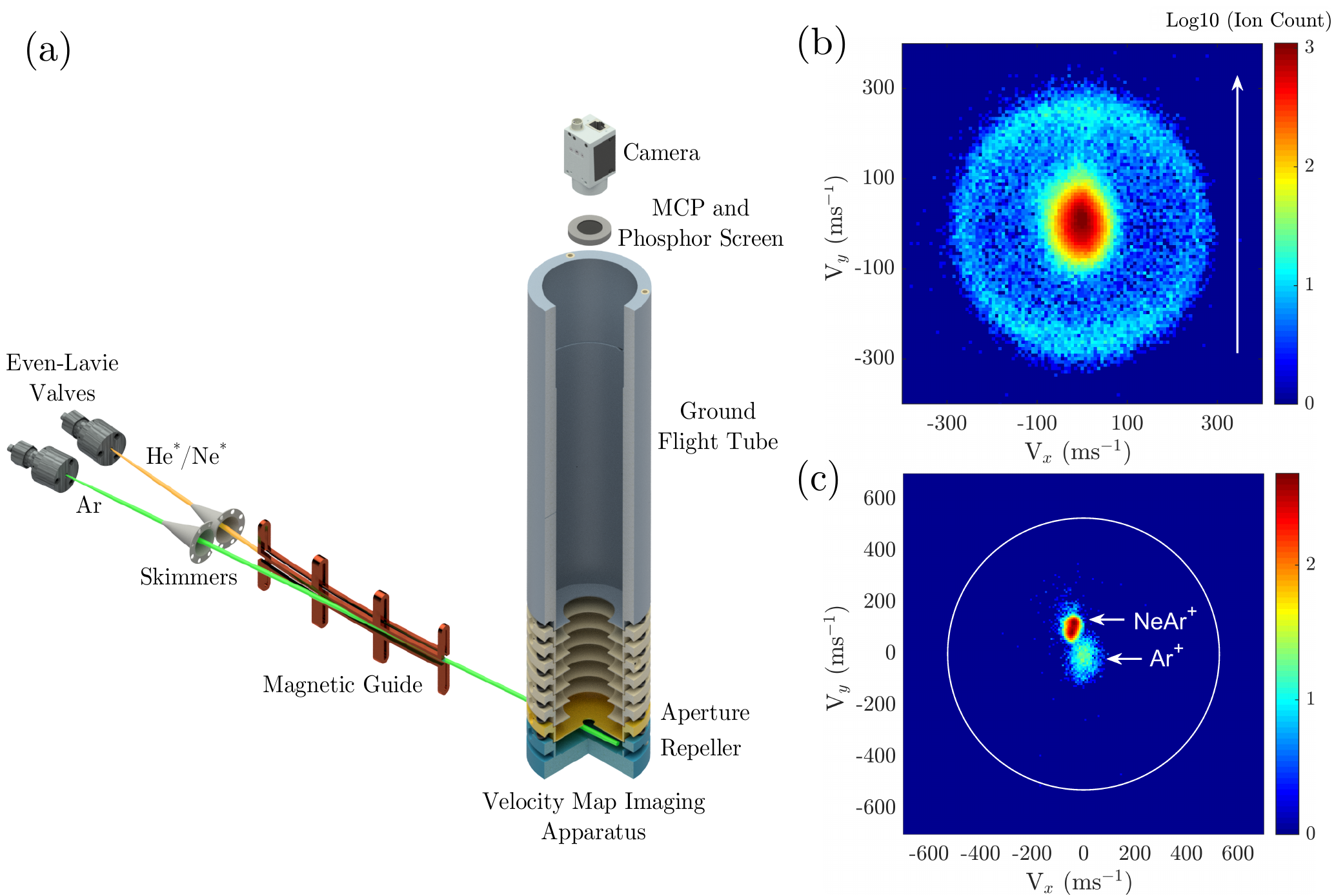}
\caption{Schematic setup of the velocity map imaging experiment (a) together with images for Ar$^+$ from He*-Ar collisions (b),  and Ar$^+$ and NeAr$^+$ from Ne*-Ar collisions (c), collected at collision energies $E/k_B=$7.8$\,$K and 14.0$\,$K, respectively. The outer ring observed for Ar$^+$ (from He*-Ar) corresponds to a kinetic energy comparable to the spin-orbit splitting, cf. Fig.~\ref{fig:lifetimes}, and 
indicates presence of FFR with lifetimes shorter than the time of flight (about 12$\,\mu$s). The absence of such a ring for Ar$^+$ (from Ne*-Ar) suggests lifetimes significantly exceeding the time of flight.}
\label{fig:vmi}
\end{figure}
Experimental evidence for the presence of short-lived FFR in HeAr$^+$ and long-lived FFR in NeAr$^+$, as predicted theoretically in Fig.~\ref{fig:lifetimes}, is provided by velocity map imaging of ions resulting from Penning and associative ionization processes occurring in a merged beam apparatus. In these experiments, as sketched in Fig.~\ref{fig:vmi}, neutral argon atoms are ionized upon collision with $^{3}S_{1}$ metastable helium or $^{3}P_{2}$ metastable neon, respectively. Ionization at large interparticle separation results in atomic products, i.e., Ar$^+$ and neutral helium or neon. In contrast, if ionization happens at short interparticle separation, molecular ions are formed~\cite{GordonNatChem18} in the three electronic states shown in Fig.~\ref{fig:lifetimes}. Molecules in the $A_2$ state may decay during the time of flight, leading to dissociation into Ar$^+$ and neutral helium or neon, respectively. The resulting gain in kinetic energy of the ions is of the order spin-orbit excitation energy of Ar$^+$, which amounts to 1431$\,$cm$^{ -1}$.  
All the ionic products are detected by a velocity map imaging setup, cf. Methods. The image presented in Fig.~\ref{fig:vmi}(b) shows the velocity distribution of argon ions produced in the Penning ionization of argon by metastable helium. The vertical and horizontal axes correspond to the argon ion velocities parallel and perpendicular to the collision axis, respectively. The central feature with a width of 35$\,$m/s is formed by Ar$^{+}$ produced directly in the Penning ionization whereas the ring of 274($\pm$25)$\,$m/s radius is formed by argon ions that are generated by the decay of $A_2$  spin-orbit excited molecular ion state of HeAr$^+$ and subsequent dissociation. The excess energy, on the order of 1431$\,$cm$^{-1}$, is distributed among the Ar$^{+}$ ion and the neutral helium fragment, leading to a recoil of Ar$^{+}$ on the order of 274($\pm$25)$\,$m/s.
In a striking difference to Fig.~\ref{fig:vmi}(b), the image shown in Fig.~\ref{fig:vmi}(c), obtained from the ionization of argon by metastable neon, does not show the outer ring. NeAr$^{+}$ molecular ions are detected in the same image and appear as a focused feature with the same width as the Ar$^{+}$ ions. The molecular ions appear at a different spot due to  deflection by a bias magnetic field applied across the VMI setup, cf. Methods. This suggests that the lifetime of spin-orbit excited molecular NeAr$^+$ ion significantly exceeds the time of flight which is on the order of several microseconds.

\begin{figure}[tb]
\centering
\includegraphics[width=\linewidth]{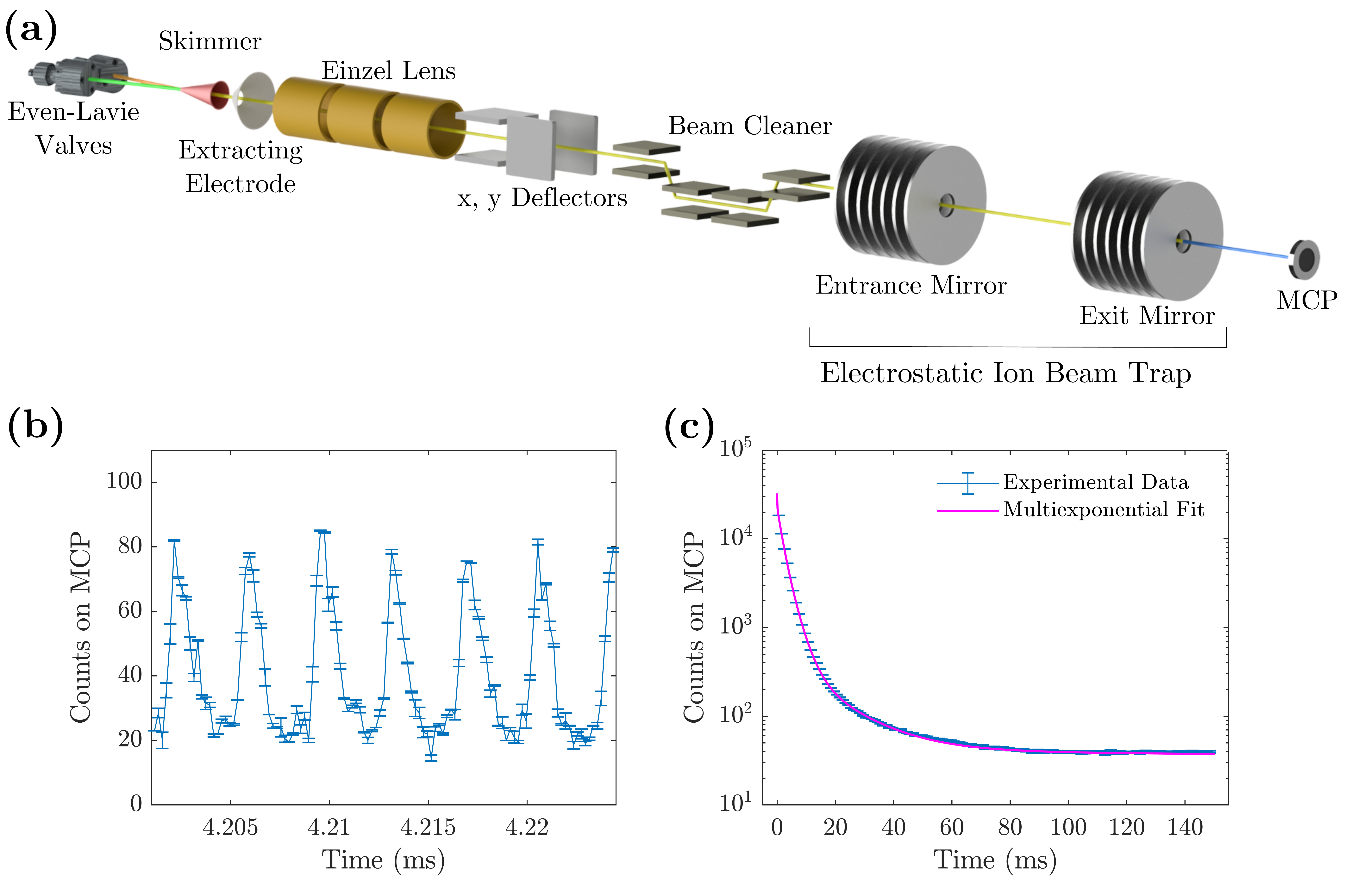}
\caption{(a) Experimental setup used to produce and trap a beam of NeAr$^+ $ and detect the neutral decay products over time. (b,c) Neutral product count versus time, after 22500 injections binned with $2\cdot 10^{-7}\,$s (b) and  $1\cdot 10^{-3}\,$s (c) together with the multi-exponential fit. }
\label{fig:lifet_exp}
\end{figure}
In order to assess the NeAr$^+$ lifetimes directly and probe, in particular, the predicted lifetime range covering several orders of magnitude, we have carried out a second experiment, cf. Fig.~\ref{fig:lifet_exp},
using two crossed molecular beams containing argon and metastable neon to generate molecular ions. 
These ions were injected into an electrostatic ion beam trap (EIBT) \cite{zajfman1997} with the apparatus shown in Fig.~\ref{fig:lifet_exp}(a) (see Methods for details), where they are trapped for several hundreds of milliseconds, during which they oscillate between two electrostatic mirrors.
Collisions with the residual gas as well as predissociation produce neutral particles that leave the trap and are detected by a micro-channel plate detector (MCP). Predissociation of molecules in the spin-orbit excited $A_2$-state, expected to populate vibrational levels $v=3,\ldots,9$ with 
$J=0.5,\ldots,35.5$
given the temperature of the associative ionization reaction, 
can thus be observed in the decay of the number of trapped ions. 
Figure~\ref{fig:lifet_exp}(b) shows the counts of neutral particles lost from the trap over time, displaying a distinct oscillation frequency which corresponds to the mass-to-charge ratio of the molecules, matching that of NeAr$^+$. Figure~\ref{fig:lifet_exp}(c) depicts the same quantity on a longer time scale together with a multi-exponential fit. We find non-exponential decay with decay times ranging from 50$\,\mu$s to 16$\,$ms, in agreement with the range of predicted lifetimes for the $A_2$ state shown in Fig.~\ref{fig:lifetimes}.


We have shown that quantum resonances may naturally be phase protected without the need for external field control. Occurrence of phase protection depends on the width, depth and equilibrium position of the bound state potential 
and can be tuned with reduced mass. Our findings provide a blueprint for protecting quantum states against undesired decay.

\section*{Methods}

\subsection*{Model of rovibrational FFR in rare gas diatomic ions}

Rovibrational FFR in rare gas diatomic ions arise due to the spin-orbit interaction coupling different non-relativistic electronic states. 
The Hamiltonian describing the rovibrational motion of a diatomic molecule including relativistic and angular couplings\cite{BusseryJCP06} can be restricted here to the lowest $^2\Sigma^+$ and $^2\Pi$ states and reads
\begin{subequations}\label{eq:model} 
\begin{eqnarray}
\label{eq:H_Hund_a}
H = T_{rad} + V_{ad} + H_{SO} + H_{rot}
\end{eqnarray}
with $T_{rad}=-\frac{\hbar^2}{2\mu}\frac{d^2}{dR^2}\mathrm{diag}(1,1,1)$ the radial kinetic energy, 
\begin{eqnarray}
\label{eq:V_Hund_a}
V_{ad} + H_{SO} = 
\begin{pmatrix}
V^{^2\Sigma^+}(R) & -\frac{\roottwo}{3}\Delta & 0 \\
-\frac{\roottwo}{3}\Delta & V^{^2\Pi}(R) + \frac{1}{3}\Delta & 0 \\
0 & 0 & V^{^2\Pi}(R) - \frac{1}{3}\Delta 
\end{pmatrix}
\end{eqnarray}
with $V^\alpha(R)$ the adiabatic non-relativistic interaction potentials  and $\Delta$ the spin-orbit interaction, and 
\begin{eqnarray}
\label{eq:Hrot}
H_{rot} = \frac{1}{2\mu R^2}
\begin{pmatrix}
J(J+1)+\frac{9}{4}+\epsilon(J+\half)&
\roottwo-\epsilon\roottwo(J+\half) & -\sqrt{2(J(J+1)-\frac{3}{4})}\\
\roottwo-\epsilon\roottwo(J+\half) & J(J+1)+\frac{5}{4}&-\sqrt{J(J+1)-\frac{3}{4}} \\
-\sqrt{2(J(J+1)-\frac{3}{4})} & -\sqrt{J(J+1)-\frac{3}{4}}& J(J+1)-\frac{3}{4}
\end{pmatrix}
\end{eqnarray}
\end{subequations}
with $J$ denoting the total molecular angular momentum quantum number and 
$\epsilon$ the spectroscopic parity, $\epsilon=\pm 1$. In Eqs.~\eqref{eq:model}, we have neglected 
the $R$-dependence of both the spin-orbit interaction and the angular couplings 
and taken $\Delta$ to be equal to the value of 
Ar$^+(^2P)$. This approximation is well justified since for HeAr$^+$ these couplings are estimated to differ from their asymptotic values by less than 1\%,\cite{GemeinCPL90} and a similar behavior is expected for NeAr$^+$.
We have also neglected the relativistic Cowan-Griffin term and the diagonal adiabatic correction, both of which are tiny. 

While all numerical results are based on Hamiltonian~\eqref{eq:model} without further approximations, we have uses the Hund's case (c) representation in our discussion which is appropriate to illuminate the rovibational structure of HeAr$^+$ and NeAr$^+$.
This implies that $J$ and $\Omega$, the body-fixed $z$-projection
of total electronic angular momentum, are good quantum numbers, with $\Omega=1/2$ for states $X$ and $A_2$ (corresponding to $^2\Sigma^+$ and $^2\Pi_{1/2}$), and $\Omega=3/2$ for state $A_1$ (corresponding to $^2\Pi_{3/2}$). 

\subsection*{Potential energy curves}

Empirical adiabatic potential curves obtained by fitting high-resolution spectroscopic data are available\cite{CarringtonJCP95} and have been used for 
the HeAr$^+$ ion, whereas
\textit{ab initio} calculations have been carried out for NeAr$^+$.
The lowest adiabatic $^2\Sigma^+$ and $^2\Pi$ electronic states of NeAr$^+$ have been obtained from 
supermolecular calculations employing a spin-unrestricted coupled cluster method
with single, double and noniterative triple excitations [UCCSD(T)], as implemented
in the Molpro package.~\cite{molpro} We have used the
correlation-consistent 
augmented aug-cc-pV6Z basis set for both atoms, and an extra set of diffuse
even-tempered functions has been added to describe the Ar atom. 

\subsection*{Calculation of lifetimes}

Resonance positions and widths have been obtained by adding a complex absorbing potential (CAP) to all three potential energy curves and subsequent diagonalisation of the resulting non-Hermitian Hamiltonian.~\cite{Riss1993}
The calculations were performed on a spatial grid extending from $3\,a_0$ to $200\,a_0$, using the Fourier method~\cite{RonnieGrid96} with $4095$ basis functions. Such a large number of grid points was necessary to properly converge the lifetimes of the narrow NeAr$^+$ resonances. 
The transmission-free CAP of Ref.~\onlinecite{Manolopoulos2002} 
with strength $E_{\mathrm{min}}=1295$~cm$^{-1}$ and absorption range $D=20\,a_0$ was used.
With this choice of parameters, the lifetimes are converged with a relative error of about $0.001$~\% for HeAr$^+$ and about $0.1$~\% for NeAr$^+$. 
$\Delta\phi$ used in Fig.~\ref{fig:lifetimes} refers to the short-range equivalent of the scattering phase shift that is defined asymptotically.

Additionally, the liftimes were calculated perturbatively using Fermi's golden rule and Hund's case (c), where 
the perturbation is given by the radial and rotational couplings. 
The Hund's case (c) representation was obtained by a unitary transformation of the full Hund's case (a) Hamiltonian. 
The FFR widths are determined as the transition rates of the $A_2$ rovibrational bound levels 
to the continua of the $X$ and $A_1$ states. Since there are no radial couplings between the $A_1$ and $A_2$ states, the $A_1$ contribution to the FFR widths, respectively lifetimes, is only minor.
Bound and energy normalized continuum states were obtained by diagonalizing the unperturbed Hamiltonian using the same grid as described above. 
The transition rates were calculated for the continuum state with energy closest to the energy of the bound level. 
For NeAr$^+$, the perturbative calculations almost exactly reproduce the lifetimes obtained non-perturbatively using the CAP and full couplings. The largest deviations occur at the maximum lifetimes, where the resonance widths become very small and overlap integrals become quite sensitive to the energy and phase of the continuum state. Nevertheless, the positions of the peaks for phase protected states are in full agreement for all rovibrational levels.
For HeAr$^+$, perturbative and non-perturbative results also agree very well except for $v=0$ and low $J$ where quantitative deviations up to a factor of four occur. Except for $v=0$ and odd spectroscopic parity, the lifetimes decrease with $J$. This monotonous dependence is in disagreement with earlier calculations that predicted an oscillatory behavior with rotational excitation.\cite{GemeinCPL90} We attribute the disagreement mainly to a different treatment of the spin-orbit interaction, with Ref.~\onlinecite{GemeinCPL90} accounting for its effect only on the potential energy curves but not on the kinetic energy.

\subsection*{Estimating the lifetime scaling with $\mu$ from the Morse oscillator approximation}

For both HeAr$^+$ and  NeAr$^+$, the $A_2$ state interaction potential for $J=0$ is well approximated around the potential minimum by the Morse oscillator,
$V(R) = D \left( e^{-2 a (R-R_0)} - 2 e^{-a (R-R_0)} \right)$
with $D$ the well depth, $a$ a constant determining the width (with smaller $a$ resulting in a wider potential well) and $R_0$ the equilibrium distance. The decay rates are given by
\begin{align*}
  \gamma_v \propto \sin\left(\delta + \phi_v(k)\right)^2 \,
  \left| \tilde\psi_v(k) \right|^2
	\;,
\end{align*}
where  $\phi_v(k) = \mathrm{arg}[\tilde\psi_v(k)]$ and $\tilde\psi_v(k)$  the Fourier transform of the Morse oscillator eigenstate. The absolute value squared $\left| \tilde\psi_v(k) \right|^2$ thus determines the minimum lifetime that can be expected. On top of a smooth variation  with $k$ (or energy) due to
$\left|\tilde\psi_v(k)\right|^2$, the decay rates $\gamma_v$ oscillate, with the roots corresponding to phase protection, cf. Eq.~\eqref{eq:condition}.
Exploiting the fact that $\tilde\psi_v(k)$ is known analytically, in terms of complex   $\Gamma$-functions,~\cite{BancewiczJPA98} we obtain  the scaling of the lifetimes, resp. decay rates, with the parameters of the potential and the reduced mass. Specifically, we find that the number of roots in $\gamma_v$ and thus the chance for phase protection increase with $D$, $R_0$, $a$, and $\mu$.
This agrees well with our observation of phase protection for NeAr$^+$ 
and its absence for HeAr$^+$, since the reduced mass and the well depth are larger for NeAr$^+$ than for HeAr$^+$, while the position of the minimum and the potential width are very similar for the two molecules.
More generally, our estimate predicts an isotope effect on predissociation lifetimes which agrees well with experimental observations for N$_2^+\,\,$ \cite{GoversJPB73} and Ne$_2^+$.\cite{GluchEPJD07}

\subsection*{Velocity Map Imaging after Merged Beam Penning Ionization }
Cold Penning ionization reactions of He$^*$-Ar and Ne$^*$-Ar are realized by the merged beam technique described by Henson et. al.\cite{HensonSci12}. The pulsed supersonic  beams of He/Ne and Ar are created by two Even-Lavie valves\cite{Even2000} positioned at a 10 degree angle. The He/Ne beam is excited to a metastable paramagnetic state by dielectric barrier discharge (DBD)\cite{Luria2009} and are deflected by a curved magnetic guide such that it is co-propagating with the Ar beam. The short opening duration of the Even-Lavie valves allow collision energies as low as 10$\,$mK $\times k_B$ to be reached.\cite{Shagam2013}

The two beams meet in the reaction region inside the velocity map imaging (VMI) detector.
The beams are skimmed by two parallel razors before the VMI detector to limit their distribution to a 1$\,$mm plane. The velocity map imaging setup consists of 8 separate plates electrodes positioned perpendicular to the beam path, followed by a 9$^{th}$ grounded electrode. These are followed by a flight tube approximately 1$\,$m long. The electrode immediately above the beam (second plate) has an aperture of 1$\,$cm in diameter for the ions to pass through, limiting the overall region from which the ions can be imaged to a 1$\,$mm by 10$\,$mm diameter cylinder. The remaining 6 electrode rings have an aperture of 40$\,$mm diameter.

Argon is ionized upon collision with He($^3S_1$)/Ne($^3P_2$), leading to Ar$^+$ as well as HeAr$^+$/NeAr$^+$ for associative ionization. The ions are then accelerated towards the micro-channel plate (MCP) detector. The MCP is followed by a phosphor screen that facilitates the imaging of the product ions by a charge couple device (CCD) camera. The center of the individual product ions is found as these are accumulated.

Since the time of ionization is not precisely known for our reaction, we must adjust the VMI method to allow for mass selectivity. To this end, we have used two  VMI operation modes: (i)  constant voltages mode  where particles with different mass-to-charge ratio  are spatially separated by an external magnetic field and (ii) a pulsed mode \cite{Mikosch2007,Mikosch2008}  that  selects particles with a certain mass-to-charge ratio by time of flight (TOF). In our experiments, the products from Penning (Ar$^{+}$) and associative ionization (HeAr$^{+}$ and NeAr$^{+}$) for the two reactions were separated using these methods.

(i) Constant voltages with bias magnetic field: In this mode, all 8 plates of VMI have constant voltages throughout the measurement. By applying a uniform magnetic field parallel to the neutral beam collision axis, the product ions get shifted on the detector according to their respective mass-to-charge ratios. The VMI resolution is not affected since the magnetic field is uniform. Figure~\ref{fig:vmi}(c) is acquired with a magnetic field of approx. 10$\,$Gauss along the entire VMI.

 (ii) Pulsed mode: This approach combines the idea of velocity mapping with TOF mass spectrometry allowing the imaging of a single mass/charge on the detector. In this method, all the plates are simultaneously pulsed to their respective imaging voltages during interaction time. The imaging pulse to the VMI electrodes follows a cleanup pulse with a period defining the `product accumulation time'. To keep the VMI resolution optimal, this timing is limited by the expansion of the products outside the imaging volume. When the imaging pulse is applied, all the ions formed during `accumulation time' fly up to the detector and get separated during their flight times, allowing TOF mass spectrometry to be performed. The MCP at the end of the flight tube is pulsed on for 50 ns selecting a single mass product for imaging. This method is used for collecting the data for Fig.~\ref{fig:vmi}(b), which shows Ar$^+$ obtained from He*-Ar collisions. Note that the actual size of the image formed on the detector is 5$\,$mm in diameter.

The velocity detected by the VMI is calibrated by creating a cold beam of Rydberg argon atoms at a range of velocities by DBD on a beam of argon. The beam is then field ionized when the VMI is pulsed and its position is detected for every velocity. The velocity of the argon beam is measured by another MCP positioned in front of the argon valve. Both the UV light from the discharge and the metastable argon atoms are detected by the MCP giving the time of flight.

\subsection*{Measurement of decay rates}
NeAr$^{+}$ ions were formed in crossed molecular beams of argon and neon, produced by Even-Lavie valves. The neon beam was excited to the metastable $^{3}P_2$ level with a dielectric barrier discharge (DBD),  with all ions generated during the excitation deflected away. The beams crossed at 12 degrees, their angles respective to the axis of the skimmer. Assuming translational beam temperatures of 1$\,$K, we obtain a collision energy of around 16$\,$meV, which corresponds to a reaction temperature of 190$\,$K. The collision products were skimmed before being accelerated to 4.2 keV, directed through several beam manipulation elements and injected into the EIBT trap. The setup is depicted in Fig.~\ref{fig:lifet_exp}(a), for further details we refer to Ref. \onlinecite{rahinov2012}. Mass-sensitive detection of the trapped particles by a pick-up electrode was used to optimize the setup parameters such as manipulation voltages and trap timing. Neutral products stemming from dissociation of the ions or neutralizing collisions with the background gas are lost from the trap and were detected on a MCP located on-axis behind the trap. NeAr$^{+}$ could be trapped with minimal contaminating ions, as confirmed by FFT analysis of the pick-up signal and the distinct oscillation observed in the neutral count, as depicted in Fig.~\ref{fig:lifet_exp}(b). The experimental data is binned in 200$\,$ns-intervals to bring out the oscillations on the scale of the NeAr${^+}$ frequency in the trap (Fig.~\ref{fig:lifet_exp}(b)), and in 1$\,$ms-intervals for showing the overall decay in Fig.~\ref{fig:lifet_exp}(c).

The measured data is fitted to a multi-exponential decay,
$f(t)=\sum_i a_i\exp(-t/\tau_i)$. The best agreement,
corresponding to a correlation coefficient of $R^2=0.995$, 
is found when including six terms in the fit, cf. Table~\ref{tab:fit},
where the longest decay time corresponds to collisions with the background gas.

\begin{table}[tbp]
\centering
\begin{tabular}[c]{|c|c|c|c|c|c|c|}
\hline
$\tau_i$ & 50.0$\,\mu$s & 1.50$\,$ms & 2.03$\,$ms & 3.44$\,$ms & 15.8$\,$ms & 1.72$\,$s \\ \hline
$\quad a_i\quad$ & $1.01\times 10^{4}$ & $7.48\times 10^{3}$ & $6.06\times 10^{3}$ & $8.31\times 10^{3}$ & 406 & 40.9 
\\\hline
\end{tabular}
\caption{Decay times $\tau_i$ with weights $a_i$ obtained when fitting the experimental data to $\sum_i a_i \exp(-t/\tau_i)$.}
\label{tab:fit}
\end{table}

\begin{acknowledgments}
CPK is grateful for a Rosi and Max Varon Visiting Professorship. 
Financial support from the German-Israeli Foundation, grant
no. 1254, is gratefully acknowledged. 
\end{acknowledgments}

\subsection*{Author contributions}

AB, WS and DMR carried out the theoretical calculations. JWR designed and YS built the VMI apparatus, and YS, PP and NB performed the VMI measurements. NJH carried out the lifetime measurements with the help of OH. EN and CPK planned and supervised the project. All authors contributed to the discussion of the results and the writing of the manuscript.

\bibliography{lifetimes}

\end{document}